\documentclass[aps,prl,twocolumn,superscriptaddress,noshowpacs,a4paper,10pt]{revtex4}
\usepackage{color}
\usepackage{amssymb}
\usepackage{graphicx}
\usepackage[pdftex]{hyperref}
\hypersetup{colorlinks=true,citecolor=blue,linkcolor=blue,urlcolor=blue}
\usepackage[all]{hypcap}

\usepackage{mathtools}
\usepackage{array}

\begin{document}

\author{Roberto Guerra}
\email{guerra@sissa.it}
\affiliation{International School for Advanced Studies (SISSA), Via Bonomea 265, 34136 Trieste, Italy}
\affiliation{CNR-IOM Democritos National Simulation Center, Via Bonomea 265, 34136 Trieste, Italy}

\author{Andrea Benassi}
\affiliation{Institute for Materials Science and Max Bergmann Center of Biomaterials, TU Dresden, 01062 Dresden, Germany}
\affiliation{Dresden Center for Computational Materials Science (DCCMS), TU Dresden, 01062 Dresden, Germany}

\author{Andrea Vanossi}
\affiliation{CNR-IOM Democritos National Simulation Center, Via Bonomea 265, 34136 Trieste, Italy}
\affiliation{International School for Advanced Studies (SISSA), Via Bonomea 265, 34136 Trieste, Italy}

\author{Ming Ma}
\author{Michael Urbakh}
\affiliation{School of Chemistry, Tel Aviv University, 69978 Tel Aviv, Israel}
\affiliation{The Sackler Center for Computational Molecular and Materials Science, Tel Aviv University, Tel Aviv 6997801, Israel.}

\title{Friction and Adhesion mediated by supramolecular host-guest complexes}

\begin{abstract}\textbf{
The adhesive and frictional response of an AFM tip connected to a substrate through supramolecular host-guest complexes is investigated by dynamic Monte Carlo simulations.
The variation of the pull-off force with the unloading rate recently observed in experiments is here unraveled by evidencing a simultaneous (progressive) break of the bonds at fast (slow) rates.
The model reveals the origin of the observed plateaus in the retraction force as a function of tip-surface distance, showing that they ensue from the tip geometrical features.
In lateral sliding, the model exhibits a wide range of dynamic behaviors ranging from smooth sliding to stick-slip at different velocities, with the average friction force determined by the characteristic formation/rupture rates of the complexes. In particular, it is shown that for some molecular complexes friction can become almost constant over a wide range of velocities.
Also, we show the possibility to exploit ageing effect through slide-hold-slide experiments, in order to infer the characteristic formation rate. Finally, our model predicts a novel ``anti-ageing'' effect which is characterized by a decrease of static friction force with the hold time. Such effect is explained in terms of enhancement of adhesion during sliding, especially observed at high driving velocities.
}\end{abstract}

\maketitle

\section{Introduction}

The field of nanotribology evolved around attempts to understand the relationship between frictional forces and microscopic properties of systems.\cite{revmod} Recent experimental and theoretical studies\cite{buda,drum,filippov,barel2010,barel2011,kim,reimann,Bennewitz1} have suggested that the observed frictional phenomena might originate from the formation and rupture of microscopic bonds (junctions) that form between surfaces in close vicinity. Furthermore, these findings indicate that stick-slip motion is connected to a collective behavior of the bonds \cite{buda,drum,filippov,barel2010,barel2011,Bennewitz1}. The formation and rupture of bonds are thermally activated processes and, thus, temperature may play an important role in the dynamics of friction at the nanoscale.\cite{filippov,barel2010,barel2011,kim,reimann}
Friction is not simply the sum of single-bond responses, but is influenced by temporal and spatial dynamics across the entire ensemble of bonds that form the frictional interface. The way how individual bonds can be averaged to yield friction response has been the focus of intense research in the past decade,\cite{revmod} however many key aspects of the friction dynamics and its relation to the kinetics of bond formation and rupture are still not well understood. One of the main difficulties in understanding and predicting frictional response is a lack of information on the nature of mediating bonds and their kinetic characteristic rates of formation and rupture.
The phenomenological models \cite{filippov,barel2010,barel2011,reimann} that successfully described velocity and temperature dependencies of friction through thermally activated rupture, formation, and strengthening of molecular bonds involved a large number of empirical parameters, which limits their insight and predictive power. In most frictional force microscopy (FFM) experiments the tip apex termination remains unknown in term of its structure and chemical nature, a fact that may severely restrict the interpretation of the data.
A significant progress in understanding microscopic mechanism of friction in terms of dynamical rupture and formation of molecular bonds can be achieved through investigations of model systems, where the tip and substrate are functionalized by assemblies of host molecules and adhesion between the contacting surfaces is caused by supramolecular host-guest interactions.\cite{Bennewitz1} In this case, in contrast to previous studies of frictions, the usage of single-molecule techniques allows more detailed insight in binding forces and rates of bond formation and rupture.\cite{evans,gomez}
First experimental study of friction and adhesion caused by cooperative rupture of supramolecular bonds\cite{Bennewitz1} discovered a remarkable difference in the dynamics of these processes: the pull-off force increased dramatically with unloading rate, while the friction force was found to be constant, over a sliding velocity range of more than three orders of magnitude. Moreover, it was suggested that different connector molecules can be bound to the same surface functionalization, allowing to control friction and adhesion using switchable connector molecules \cite{Bennewitz2} or to adapt  to surface roughness by varying the lengths of the connectors.

In this article we present results of simulations of dynamics of adhesion and friction caused by cooperative rupture of supramolecular bonds. Particular attention is given to the effect of a non-equal load sharing between the bonds, which results from the curved shape of the AFM tip, on the dynamics of adhesion and friction. The rate dependence of the pull-off force and velocity dependence of friction have been calculated in a broad range of rates of formation and rupture of molecular bonds, and different dynamical regimes have been discovered.

\begin{figure}[!b]
\centering
\includegraphics[width=0.8\columnwidth]{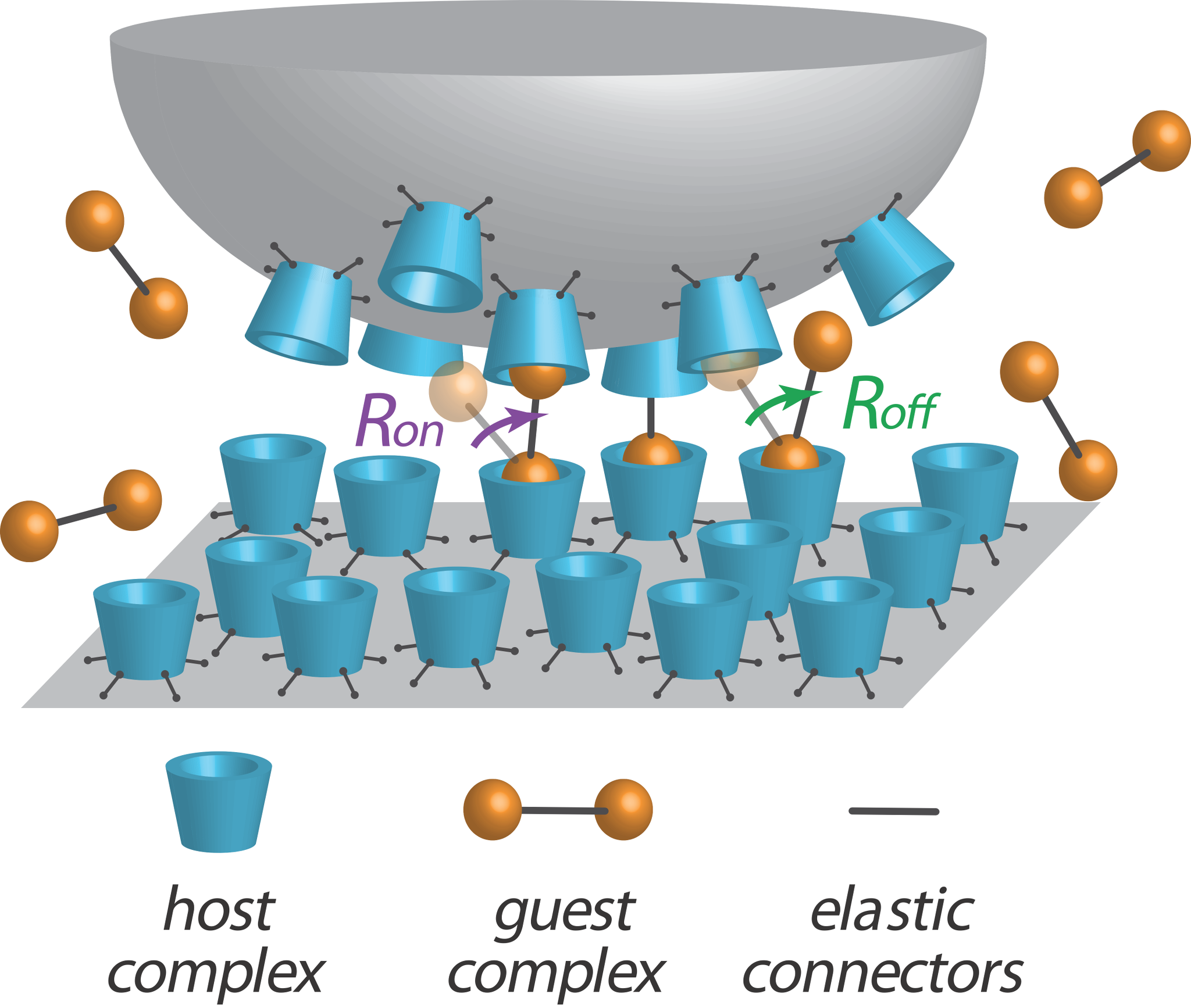}
\caption{Sketch of the system.}
\label{figure1}
\end{figure}

\section{Model}

\begin{figure}[!t]
\centering
\includegraphics[width=0.8\columnwidth]{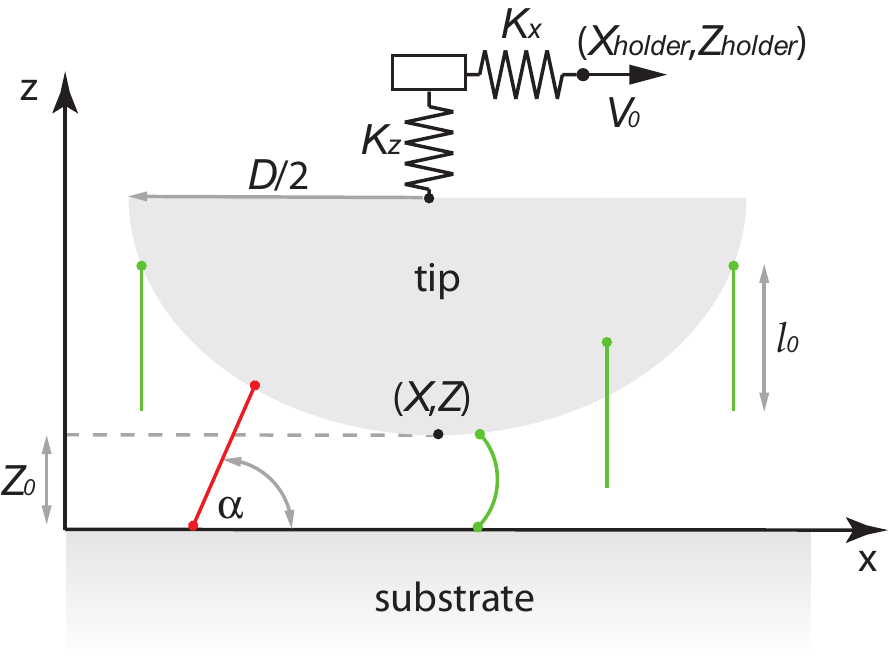}
\caption{Sketch of the adopted model. The rubber bands exerting a force are red colored, the compressed or detached ones, exerting no force, are green colored.}
\label{figure2}
\end{figure}

We employ a model in which the host complexes, represented by ideal rubber bands with spring-constant $k$ and rest length $l_0$, are attached with homogeneous density to the surface of a hemispherical 3D tip, and to the counter-surface substrate (see Fig.~\ref{figure2}).
The tip, with mass $M$, diameter $D$, and coordinates $(X,Z)$, is connected to the cantilever holder at $(X_{holder},Z_{holder})$ through springs of lateral and normal stiffnesses $K_x$ and $K_z$, respectively. The tip is then subject to a viscous damping $\gamma$, accounting for water environment. Tip and holder are initially placed at $X_{holder}$\,=\,$X$\,=\,0, $Z_{holder}$\,=\,$Z$\,=\,$Z_0$. The resulting equations of motion are

\begin{subequations}
\begin{align}
\begin{split}\label{eq.motionX}
M\ddot{X} & = K_x ( X_{holder} - X ) -M\gamma\dot{X} + \\
          & ~~~~~~~~~~~~ + \sum_{i=1}^N -k\theta(l_i-l_0)\cos(\alpha_i),
\end{split}\\
\begin{split}\label{eq.motionZ}
M\ddot{Z} & = K_z ( Z_{holder} - Z ) -M\gamma\dot{Z} + A_{sub}Z^{-7} \\
          & ~~~~~~~~~~~~ + \sum_{i=1}^N -k\theta(l_i-l_0)\sin(\alpha_i),
\end{split}
\end{align}\label{eq.motion}
\end{subequations}

\noindent where $N(t)$ connected complexes (up to $N_{max}$), forming angles $\alpha_i$ with the substrate, have elongations $(l_i-l_0)$ modulated by the Heaviside step-function $\theta$ in order to suppress repulsive contributions, that are negligible within the rubber-band assumption. The term $A_{sub}Z^{-7}$ describes a short-range repulsion between the tip and substrate.\cite{israelachvili}

The bonds form vertically, and follow the motion of the tip along $X$ direction; thus no forces along the transverse $Y$ direction is generated. The complexes bind or unbind with thermally-activated rates $R_{on}$ and $R_{off}$ expressed by

\begin{equation}
\begin{aligned}
R_{on}  & = R_{on}^0\exp[-\theta(l-l_0)\frac{1}{2}k(l-l_0)^2/k_BT], \\
R_{off} & = R_{off}^0\exp[\theta(l-l_0)k(l-l_0)\delta_l/k_BT],
\end{aligned}\label{eq.ron_roff}
\end{equation}

\noindent in which we have introduced the characteristic attempt rates $R_{on}^0$ and $R_{off}^0$, the rupture barrier length $\delta_l$, and the thermal energy $k_BT$.
It should be noted that, because of the curved shape of the AFM tip, the rates $R_{on}$ and $R_{off}$ depend on the position of the attached bond with respect to the tip apex. This configuration leads to a non-equal load sharing between the bonds.
We also note that due to the nature of the problem and the symmetry of such a geometrical configuration, the addition of a lateral direction $Y$ in the model would make the computational efforts just more demanding without adding any valuable physical insight.

The equations of motion are integrated using a velocity-Verlet algorithm with a time step $\Delta t$\,=\,1\,$\mu$s, and formation/rupture of the complexes is stochastically governed according to the probabilities $P_{on}$\,=\,$R_{on}$$\Delta t$ and $P_{off}$\,=\,$R_{off}$$\Delta t$ at each time step.

\noindent If not differently stated, the employed parameters are: $N_{max}$\,=\,100, $k$\,=\,0.16\,N/m, $l_0$\,=\,10\,nm, $\delta_l$\,=\,0.1\,nm, $Z_0$\,=\,0.1\,nm, $k_BT$\,=\,25.85\,meV, $K_x$\,=\,50\,N/m, $K_z$\,=\,0.23\,N/m, $\gamma$\,=\,13.3\,ms$^{-1}$, $M$\,=\,1.15$\times$10$^{-9}$\,kg, $D$\,=\,20\,nm. Reference values have been taken from typical AFM setup and from recent experimental data we intend to reproduce within our modeling approach.\cite{Bennewitz1}

\section{Pull-off}

\begin{figure}[!b]
\includegraphics[width=\columnwidth]{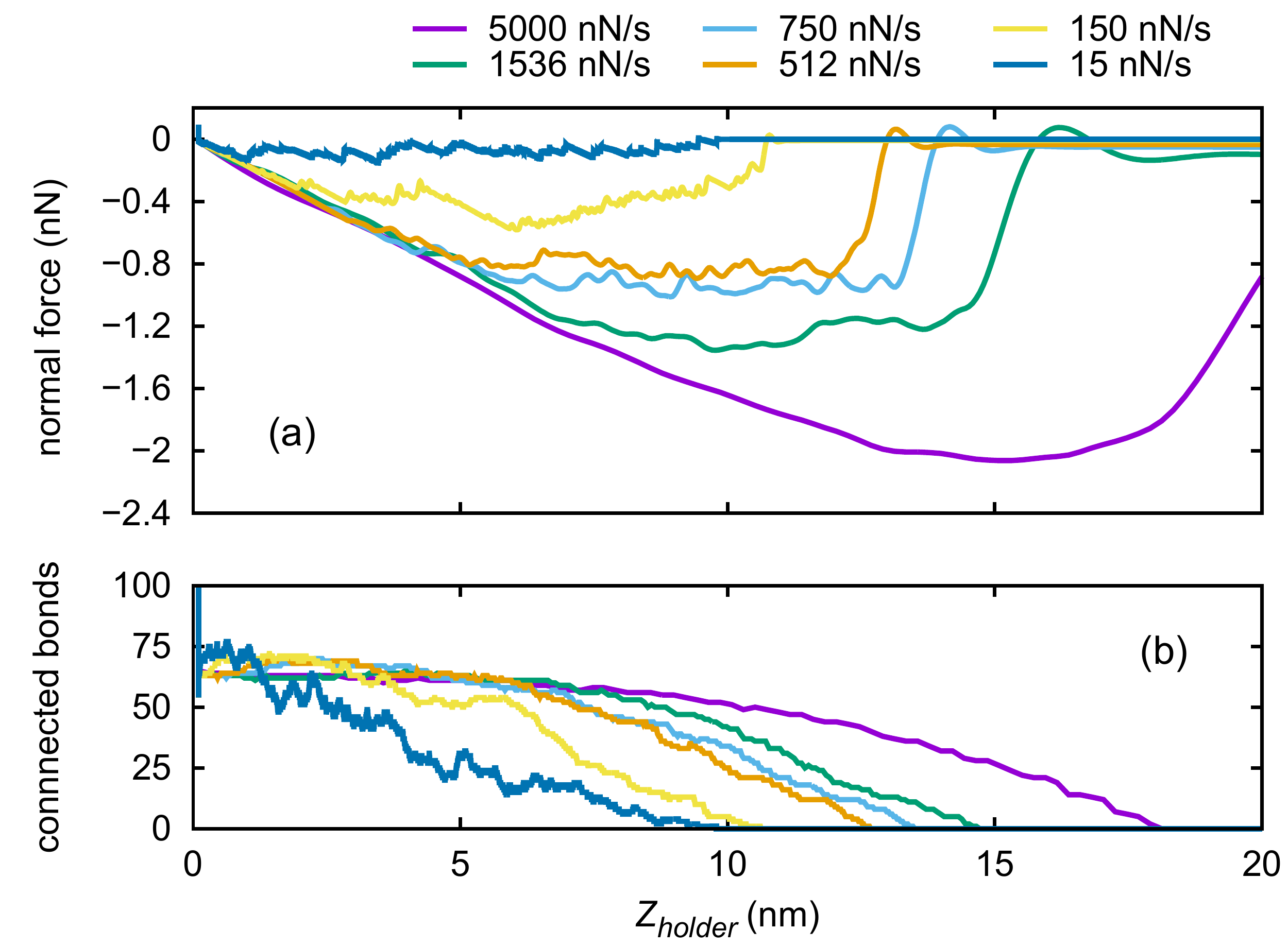}\\
\includegraphics[width=\columnwidth]{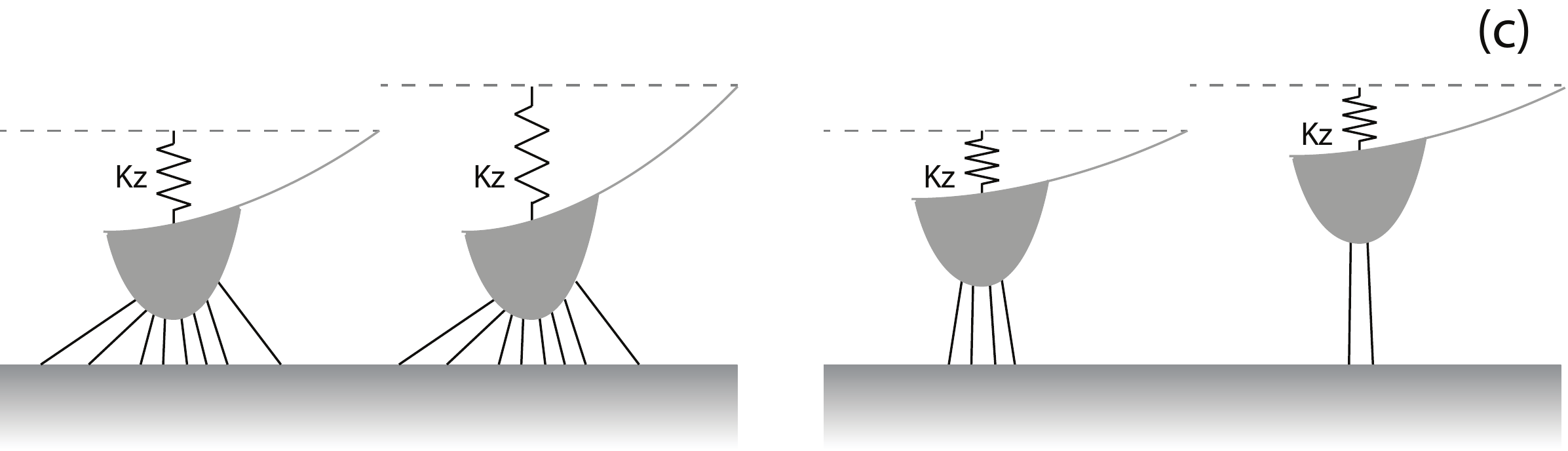}
\caption{(a) Normal force acting on a hemispherical tip for different unloading rates as a function of the cantilever holder position, with $R_{on}^0$\,=\,$R_{off}^0$\,=\,0.1\,kHz; (b) instantaneous number of connected complexes. (c) Sketch of the two different stretching scenarios during the vertical pulling: fast rate (left), slow rate(right)}
\label{figure3}
\end{figure}

The friction measurements alone do not allow to establish a microscopic mechanism of energy dissipation. In order to gain better understanding of the kinetics of cooperative rupture and formation of supramolecular bonds, the mechanical response of the system upon retraction from the surface has been measured.\cite{Bennewitz1} To understand the experimentally observed behavior, we have performed numerical simulations of the normal force as a function of the distance between the cantilever holder and the surfaces, $Z_{holder}$.

The normal force is determined by fixing the vertical velocity of the cantilever holder, $\dot{Z}_{holder}$, corresponding to a unloading rate $\dot{Z}_{holder}\times K_z$, and by measuring the force exerted on the tip, i.e. $K_z(Z_{holder}-Z)$. The normal force trend as a function of the tip height $Z_{holder}$ for different unloading rates is reported in Figure~\ref{figure3}(a).
Before the complete detachment of the complexes, and the consequent release of the tip, we can identify two different regimes: a linear (negative) increase of the force with increasing $Z_{holder}$, and a plateau region with normal force almost independent of $Z_{holder}$. The linear growth of the force becomes particularly noticeable for sufficiently high unloading rates, for which the number of connected complexes is very slowly vaying in time until complete detachment occurs; here, the detailed unbinding dynamics is unable to respond fast enough during the cantilever retraction [see Figure~\ref{figure3}(b)].
This almost constant number of bonds is that at the initial equilibrium, when $Z_{holder}$\,=\,$Z$\,=\,$Z_0$, and is solely determined by the ratio $R_{on}/R_{off}$.
As depicted in Figure~\ref{figure3}(c), when the cantilever holder gets displaced the system must equilibrate the tension among a parallel of many complexes connected in series with the cantilever spring, so that the latter must burden most of the overall stretching. Therefore, the slope of the linear increment of the normal force is mostly determined by the cantilever normal stiffness $K_z$.
The plateau region in the retraction force traces is a characteristic feature of the discussed experiments with supramolecular bonds, and as far as we know this effect was not observed in previous force measurements.
As shown in Fig.~\ref{figure3}(b) in the plateau region the number of connected complexes decreases with higher $Z_{holder}$, thus diminishing the effective stiffness of the complexes in parallel, and compensating for the stress of the cantilever spring.
The rupture starts at the periphery of the tip and propagates towards its apex.
The progressive unbinding of the complexes gives rise to a saw-tooth-shaped fine structure of the normal force in plateaus, as recently observed in experiments and exploited to determine the effective stiffness of a single complex.\cite{Bennewitz1}

\subsection{Rate dependence}

\begin{figure}[!t]
\includegraphics[width=\columnwidth]{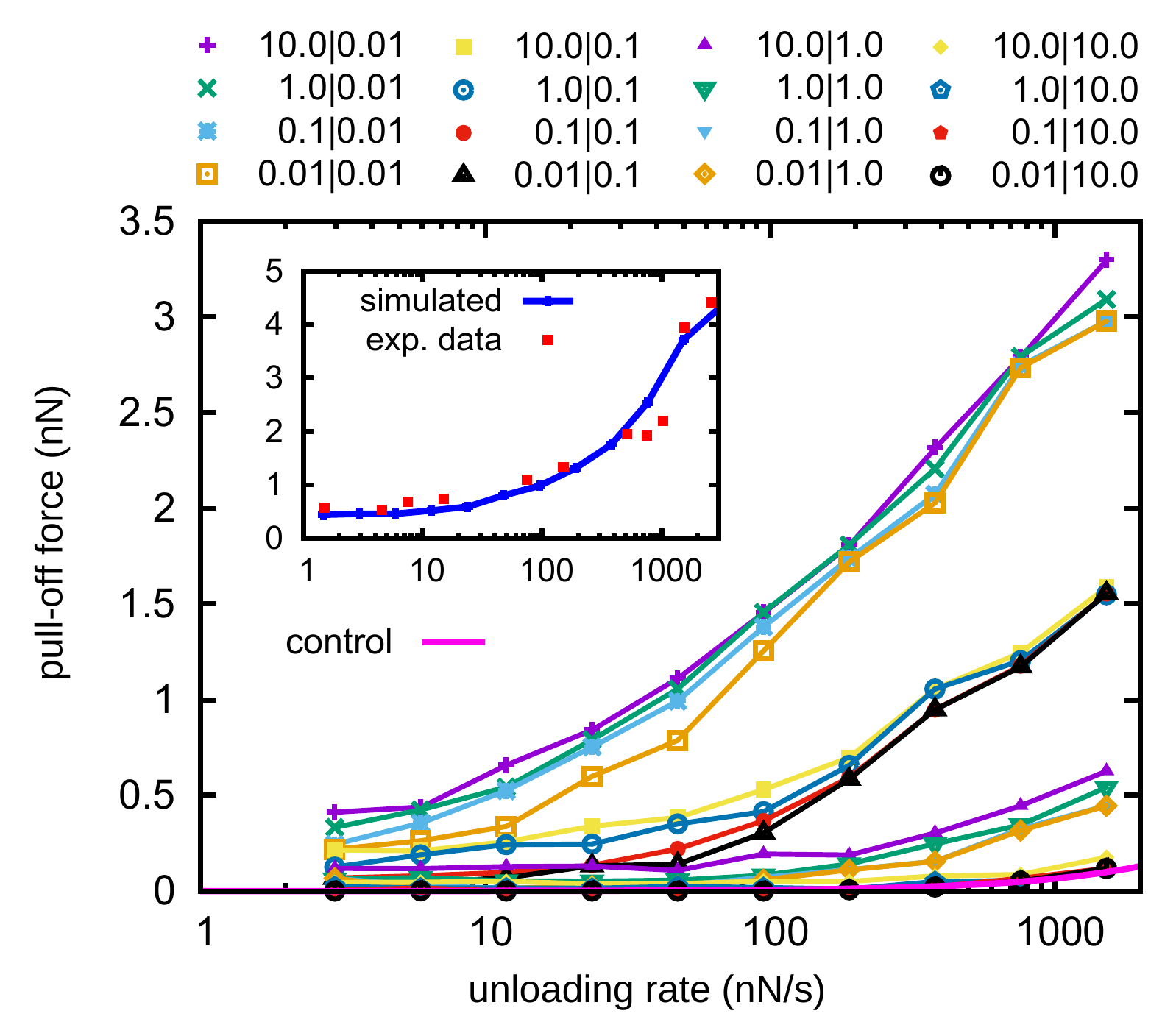}
\caption{ Pull-off force (maximum normal force of Fig.\ref{figure3}) as a function of the unloading rate for different $R_{on}^0{\mid}R_{off}^0$ pairs as reported in legend (in kHz). Control curve is with free tip, i.e.\ subject only to viscous damping. (inset) Best fit of experimental data from Ref.~\citenum{Bennewitz1}, obtained with $R_{on}^0$\,$=$\,100\,kHz, $R_{off}^0$\,$=$\,25\,Hz, $\delta_l$\,$=$\,0.22\,nm, $k$\,$=$\,0.0036\,N/m. }
\label{figure4}
\end{figure}

As anticipated above, the maximum normal force -- the pull-off force (or rupture force) -- is therefore strongly dependent on the unloading rate, and on its relation with the binding/unbinding rates in Eq.~\ref{eq.ron_roff}.
To investigate this aspect, we have reported in Figure~\ref{figure4} the pull-off forces calculated for a hemispherical tip, as a function of the unloading rate for a set of $R_{off}^0$,$R_{on}^0$ values.

In agreement with experimental observations,\cite{Bennewitz1} our results exhibit two distinct regimes: (i) at low unloading rates the rupture force is approximately constant, (ii) with increasing unloading rates $U_r$, the pull-off force grows as $U_r^{0.34}$. A comparison of our simulated rate-dependence of pull-off force with the experimental observations of Ref.~\citenum{Bennewitz1} is provided in Fig.~\ref{figure4} inset.

For increasing unbinding rates $R^0_{off}$, the transition separating these two regimes shifts at higher unloading rates. Regime (i) is usually attributed to a quasi-equilibrium state in which the number of bonds is maintained approximately constant by a fast rebinding of broken bonds.\cite{friddle} In our system the origin of this regime is different, it results from the progressive unbinding of bonds.

Besides, since 1/$R_{off}$ represents a characteristic time for bond rupture, more bonds are connected at lower $R_{off}^0$ values, resulting in larger pull-off forces. Instead, $R_{on}^0$ produces negligible effects in its considered range, due to the fact that the formation of bonds with large elongation demands energies much larger than $k_BT$, thus requiring huge $R_{on}^0$ prefactors in Eq.~\ref{eq.ron_roff} to compensate over the exponential term.
Here, we note that thermal energy is overcome already at elongations of $\sqrt{2 k_BT/k}$\,$\simeq$\,0.23 nm and $k_BT/k\delta_l$\,$\simeq$\,0.26 nm for bond formation and bond rupture, respectively. Beyond these elongations, very large (small) $R_{on}^0$ ($R_{off}^0$) prefactors are required to compensate the exponential terms in Eq.~\ref{eq.ron_roff}. Thus, within the considered range for $R_{on}^0$, the formation of a bond becomes impossible already at modest elongations, so that the bonds will never reform after rupture at the typical unbinding time-scales.
However, when a bond is intact, the elongation before rupture is (roughly speaking) determined by the rupture time 1/$R_{off}$ times the cantilever holder velocity. Clearly, larger unloading rates correspond to larger rupture lengths. This is why, as shown in Fig.~\ref{figure4}, in the investigated range of parameters the pull-off forces much increase at lower $R_{off}^0$, while $R_{on}^0$ acts as a mild pull-off shifter.

\subsection{Tip-geometry effects}

\begin{figure}[!t]
\includegraphics[width=\columnwidth]{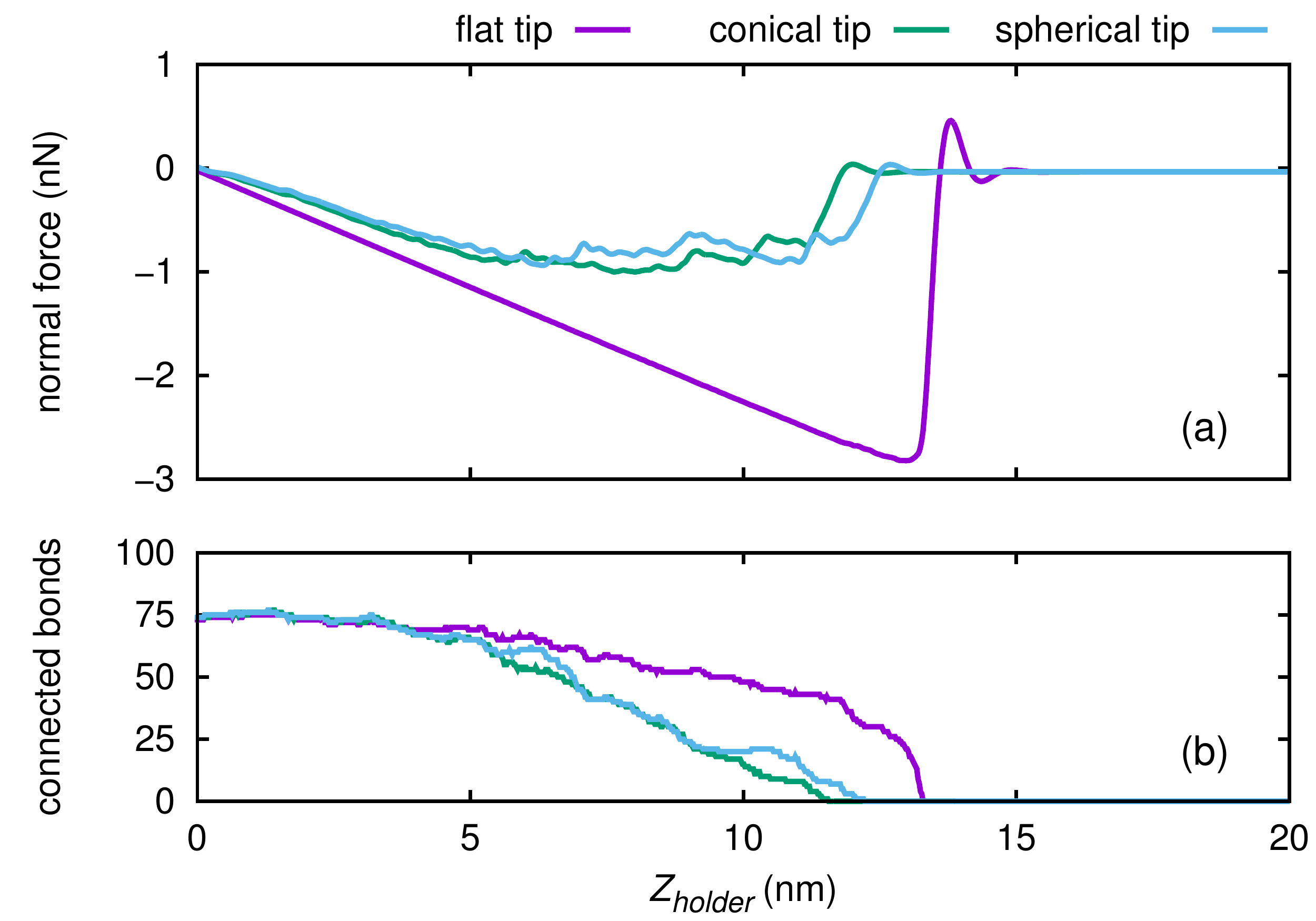}
\caption{(a) Normal force as a function of the cantilever holder position for different tip geometries (unloading rate = 512\,nN/s, $R_{on}^0$\,=\,$R_{off}^0$\,=\,0.1\,kHz); (b) instantaneous number of connected complexes.}
\label{figure5}
\end{figure}

Under the condition of identical bonds, having the same rest lengths and the same binding/unbinding probabilities, which is characteristic for a blind tip with an extended flat surface, one might expect an almost simultaneous rupture of all the bonds when the applied force exceeds a critical value, rather than a one-by-one rupture as occurring during the normal force plateaus. However, in a common experimental setup, because of the curved shape of the AFM tip, some bonds are more stretched than others, thus giving rise to a non-uniform distribution of unbinding probabilities in the ensemble of bonds.
To explore this aspect we have performed simulations with different tip shapes. In particular, in Figure~\ref{figure5} we report normal force curves obtained for flat, hemispherical, and conical tips. Clearly, both spherically- and conically-shaped tips produce similar curves, with well pronounced plateaus of the normal force extending for more than 5\,nm. Conversely, in the case of a flat tip no plateu forms, supporting the idea that the one-by-one rupture occurring at the plateaus is intimately connected to a non-uniform distribution of bond elongations induced by the tip geometry. The bonds connected at the periphery of the tip become more stretched, and they broke first.

\section{Friction}

\begin{figure*}[!t]
\centering
\includegraphics[width=8.75cm]{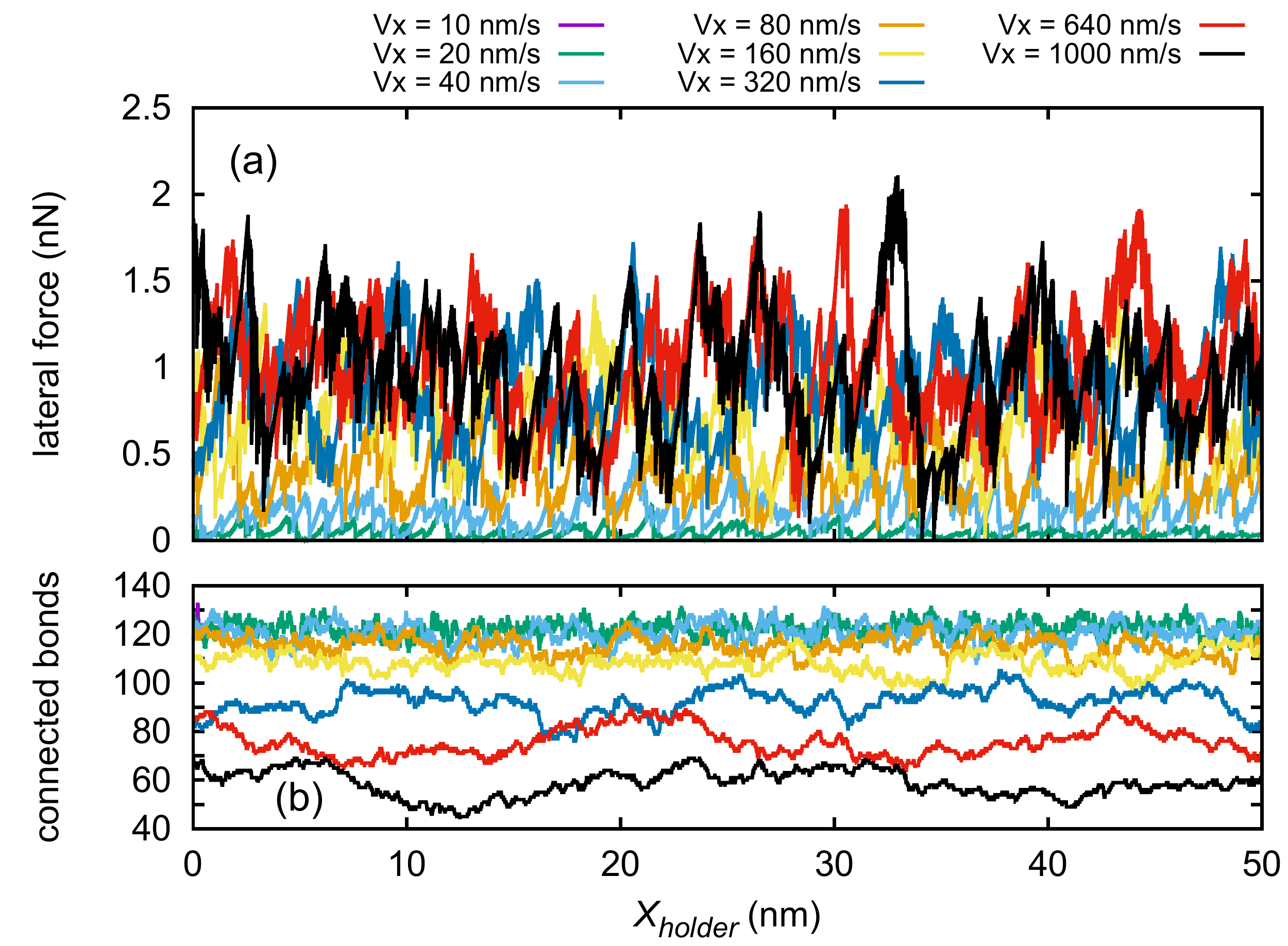}\includegraphics[width=8.75cm]{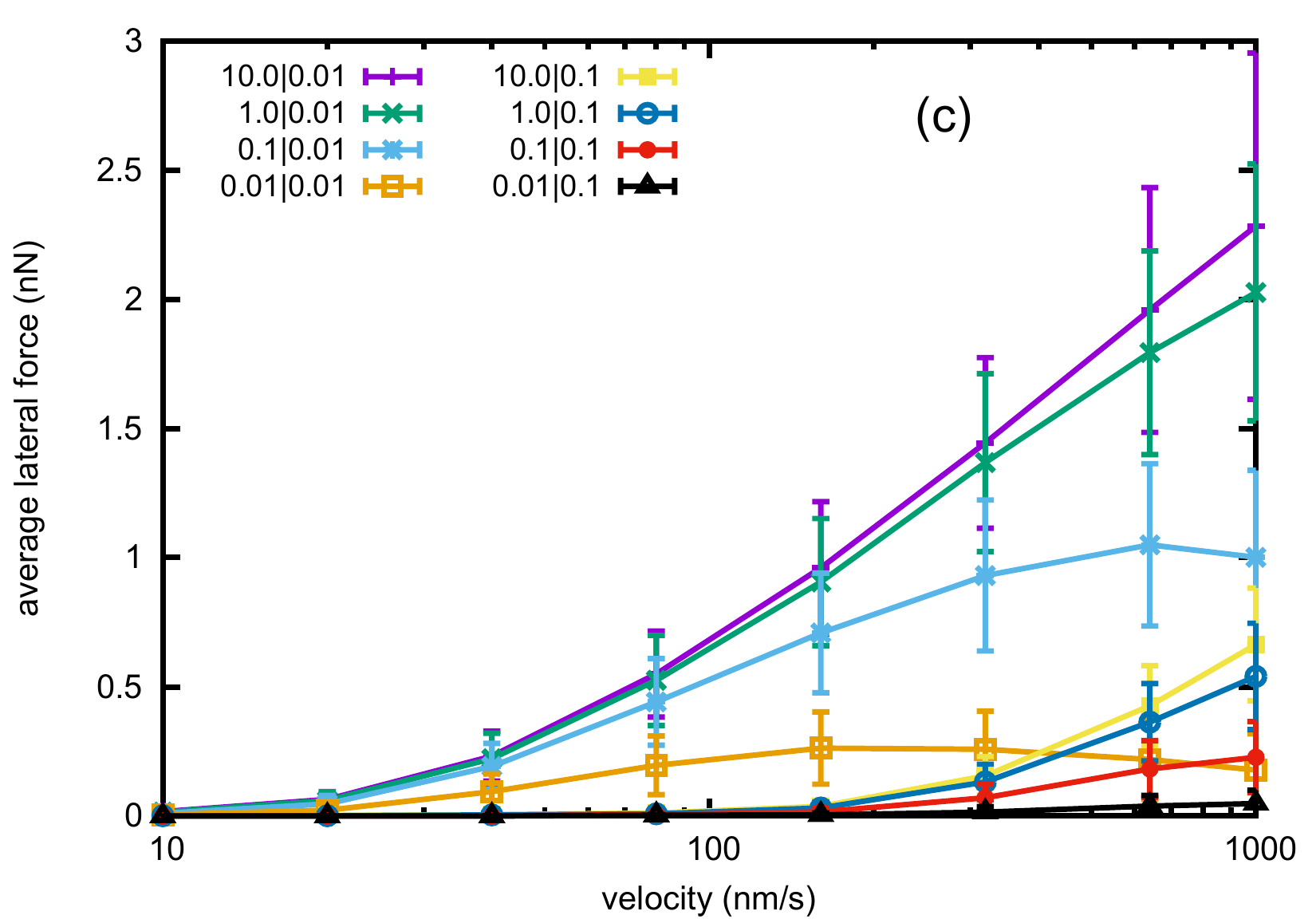}\\
\includegraphics[width=10cm]{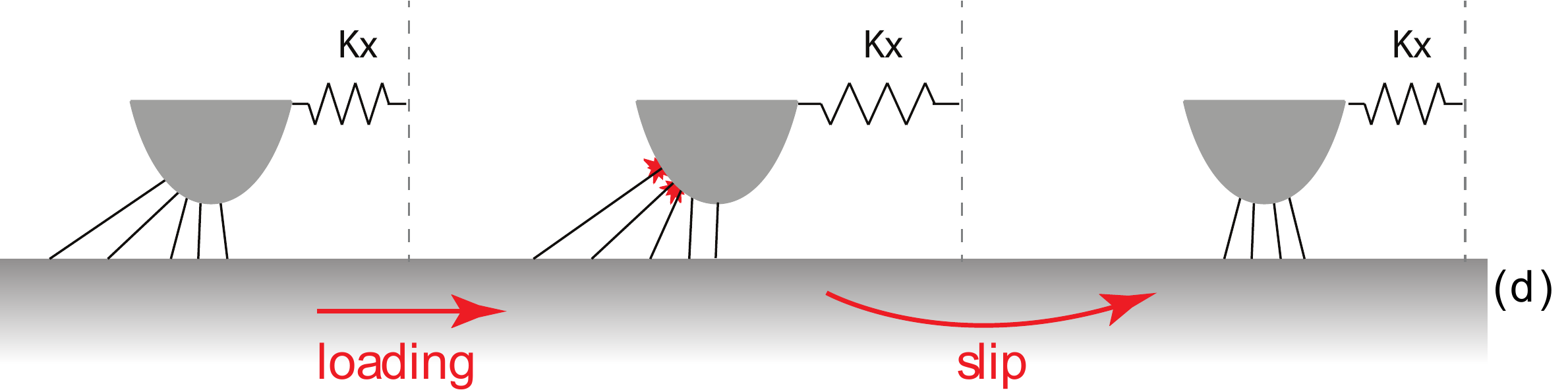}
\caption{(a) Friction force for different sliding velocities as a function of the cantilever holder position ($R_{on}^0$\,= 0.1\,kHz, $R_{off}^0$\,=\,0.01\,kHz) and (b) instantaneous number of connected complexes; (c) average lateral force as a function of the sliding velocities for different binding/unbinding rates. $R_{on}^0{\mid}R_{off}^0$ parameters for each curve are reported in the legend in kHz. Error bars report standard deviation over 200\,nm sliding; (d) Sketch of the bond breaking mechanism leading to a chaotic stick-slip motion. }
\label{figure6}
\end{figure*}

In sliding experiments, the friction force is determined by fixing the height, $Z_{holder}$, and the driving velocity, $\dot{X}_{holder}$, of the cantilever holder, and measuring its lateral torsion, $K_x(X_{holder}-X)$. Figure~\ref{figure6}(a) reports the lateral force traces as a function of the cantilever position for different sliding velocities.
We notice two opposite limiting behaviors: a smooth almost-constant force at small sliding velocity, and a chaotic stick-slip regime, with strongly increased average value (friction), at fast sliding.
As reported in Figure~\ref{figure6}(b), the average number of connected bonds during sliding is roughly constant at low velocity, revealing a steady-state regime in which the complexes bind and unbind at rates faster than the characteristic sliding frequency $1/\tau = (k\dot{X}_{holder}\delta_l/k_BT)$.
In this quasi-equilibrium regime the time-averaged friction force grows with the driving velocity [see Fig.~\ref{figure6}(c)].
On the contrary, at large velocities, we find pronounced sudden drops in the force traces of Fig.~\ref{figure6}(a), revealing the occurrence of avalanches of breaking contacts. Similar to the pull-off case, the latter regime onsets when the characteristic sliding frequency is higher than the rate of the natural unbinding timescale, thus increasing the lateral force until some bonds induce the collective rupture that leads to an abrupt slip of the tip. This mechanism is sketched in Figure~\ref{figure6}(d).

\subsection{Role of velocity and binding/unbinding rates}

By plotting the average lateral (friction) force as a function of the driving velocity for different binding/unbinding rates [Fig.~\ref{figure6}(c)], we observe a multiplicity of possible distinctive regimes, with increasing or decreasing friction as a function of velocity, and, in some case, the occurrence of a blunt maximum.
To understand such scenario we compare the characteristic sliding time with the two microscopic characteristic times of our model, namely $1/R^0_{on}$, the minimum bond formation time, and $1/R^0_{off}$, the maximum bond rupture time.
During sliding, the bonds are continuously broken and reformed in an unstressed state and the friction force can be estimated as $k\dot{X}_{holder} N_{eq}/R^0_{off}$,\cite{filippov} where $N_{eq}$ is the number of intact bonds in equilibrium.

A characteristic feature of low sliding velocity regime is the essential absence of correlation between individual rupture events, which is a manifestation of thermal bond dissociation. With increase of the speed, processes of spontaneous (thermal) and shear-induced bond dissociation start to compete, and we observe an erratic stick-slip dynamics.

The stick-slip motion becomes more pronounced at larger velocities, $\tau$\,$<$\,$1/R^0_{off}$, where the rupture is completely determined by the effect of shear stress on the height of the unbinding potential barriers. In contrast to the low-velocity sliding regime where rupture events are uncorrelated, the stick-slip motion is characterized by a cooperative behavior of bond subsystem. 
It should be noted that this high friction regime occurs only under the condition that bonds can quickly reattach, that happens when $\tau$\,$\geq$\,$1/R_{on}$. Instead, when the sliding velocity is too high ($\tau \ll 1/R_{on}$), only a small number of bonds survives, and the friction force is reduced and becomes mainly determined by the contribution coming from the viscous damping of the aqueous solution.

It is worth to note that, differently from the previous simulations of friction in terms of rupture and rebinding of interfacial junctions,\cite{filippov,barel2010,barel2012} Figs.~\ref{figure6}(a),(b) do not show a clear correlation between the stick-slip oscillations of friction force and the variation of number of attached bonds with the cantilever displacement. The important difference of the present system from those considered before is that the equilibrium length of the supramolecular bonds is significantly larger than the characteristic distance between the tip apex and the substrate. As a result, during the frictional motion most of attached bonds are in the unstretched state, and only a small percentage of them gets ruptured at the stick-to-slip transition.

Our simulations demonstrate that, by changing the ratio $R^0_{on}/R^0_{off}$ we can tune the position of the maximum of the friction force as a function of velocity [see Figure~\ref{figure6}(c)], and even push it to range of low velocities, or of very large velocities being left with a monotonically increasing trend only. Besides, there exist values of the ratio $R^0_{on}/R^0_{off}$ which lead to regimes of quasi-constant friction over a broad velocity range, reproducing pretty similar trends of recent experimental data.\cite{Bennewitz1}

The above-described variety of frictional trends has been experimentally observed through the adoption of different molecular sets.\cite{Bennewitz3}

\begin{figure*}[!hbt]
\includegraphics[width=\columnwidth]{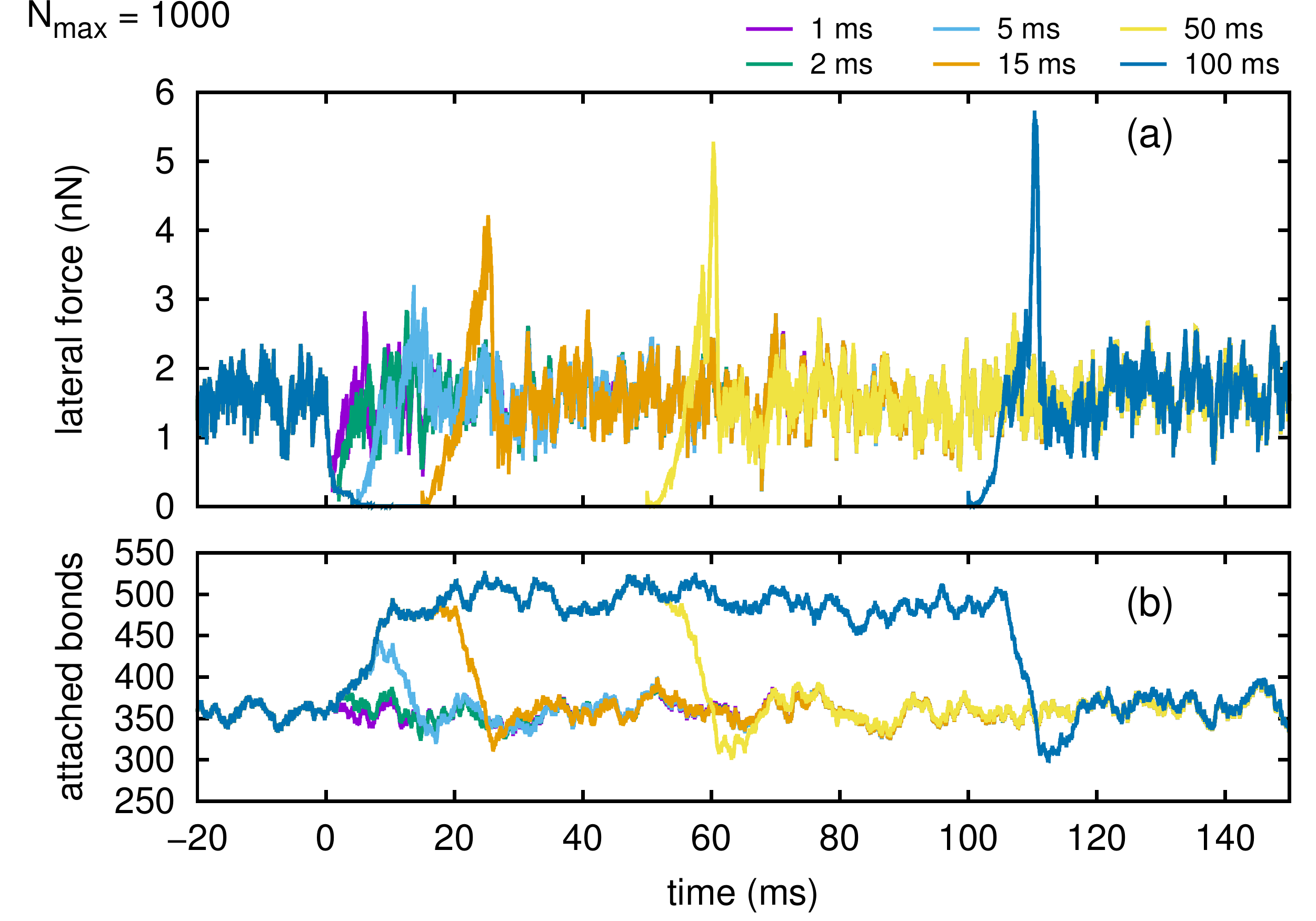}
\includegraphics[width=\columnwidth]{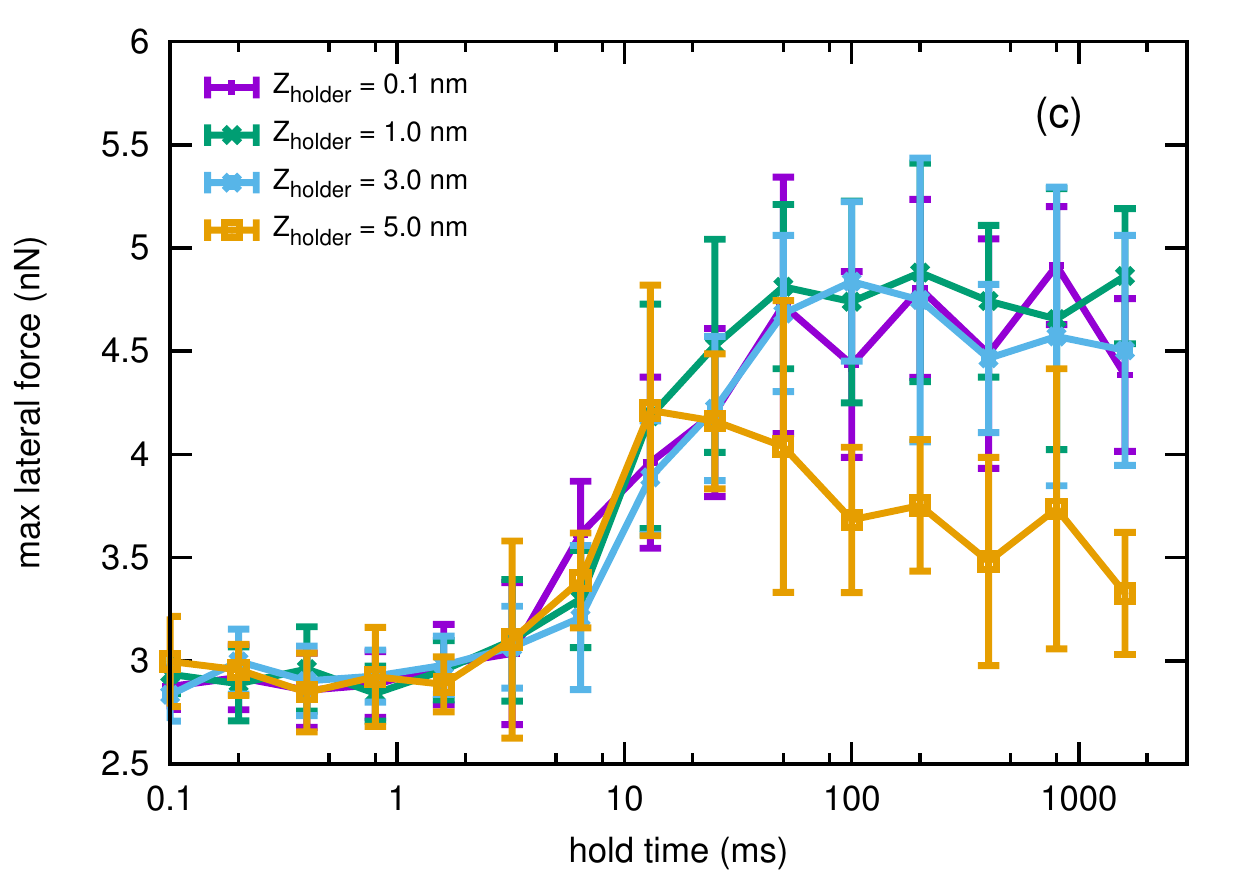}
\caption{Slide-hold-slide simulations for different hold times ($N_{max}$\,=\,1000, $V_x$ = 1000\,nm/s, $R_{on}^0$\,=\,$R_{off}^0$\,=\,0.1\,kHz): (a) Friction force as a function of time (hold starts at $t$\,=\,0); (b) instantaneous number of connected complexes; (c) Maximum lateral force as a function of the hold time for different $Z_{holder}$ values (average of 10 runs, with error bars reporting standard deviation).
}
\label{figure7}
\end{figure*}

\subsection{Ageing}

As shown by previous studies,\cite{li,barel2012} slide-hold-slide (SHS) experiments are able to access the characteristic bonding time $1/R_{on}$ of a certain molecular complex.
In this kind of experiments the sliding motion of the tip is suddenly stopped and restarted after a certain holding time $t^{\star}$. These measures are usually carried out in order to estimate the ageing of the tip-substrate junctions, i.e. to which extent the static friction is increased by the formation of bonds or capillary bridges during the rest time.
The friction force measured in our SHS simulations is reported in Figure~\ref{figure7}(a) for different holding times. As expected, the number of intact bonds increases with $t^{\star}$, as well as the resulting force maximum occurring upon resuming sliding [Fig.~\ref{figure7}(b)].
However, when $t^{\star} \gg 1/R_{on}$ the number of connected complexes can reach equilibrium, and the maximum force saturates [Fig.~\ref{figure7}(c)].
The above argument can provide a way to directly estimate $R_{on}$ by SHS experiments: in our case the maximum lateral force starts to increase at $t^{\star}$\,$\sim$\,10\,ms, in agreement with the used $R_{on}$ value of 0.1\,kHz.

In Fig.~\ref{figure7}(c) we report the maximum lateral force (static friction) also calculated for different distances between the holder and the surface, $Z_{holder}$, from -1.0\,nm to 5.0\,nm.
Here negative values of $Z_{holder}$ correspond to the regime of repulsive tip-surface interaction, while for $Z_{holder}$\,=\,5.0\,nm there is a strong attraction between the tip and the surface.
For a resting tip, while at zero or negative $Z_{holder}$ values the bonds can easily form across the whole tip surface, at larger $Z_{holder}$ the bonds form preferentially at the tip apex.
However, during lateral motion, the tip is subject to an attractive force, acting in the normal direction, that arises due to bond stretching as described by the last term of Eq.~\ref{eq.motionZ}. This normal component of attractive force grows with the tip velocity. Therefore, even for large $Z_{holder}$, during sliding the tip is attracted toward the surface, and its distance from the surface becomes very small, entering the repulsive regime.
Supramolecular bonds newly-formed during the hold time do not produce an attractive force acting in normal direction, since typical tip-substrate distances are smaller than the equilibrium bond lengths (this is a characteristic feature of long supramolecular bonds). Thus, the absence of the bond-induced attraction between the tip and substrate may lead to an increase of the tip-substrate distance during the hold time, and to the corresponding reduction of the number of attached bonds. This counterintuitive ``anti-aging'' effect becomes evident for large values of $Z_{holder}$, as shown in Fig.~\ref{figure7}(c) for $Z_{holder}$\,=\,5.0\,nm. Such effect has not been observed so far experimentally, since repulsive regime is usually adopted in experiments as reference to assure a tip-substrate contact.

\section{Conclusions}

To summarize, a model for the description of dynamics of friction and adhesion caused by cooperative rupture and formation of supramolecular bonds has been proposed and investigated. We found that a non-equal load sharing between the intact bonds, which results from the curved shape of the AFM tip, strongly influences the dynamics of friction and adhesion. In particular, this effect leads to the plateaus in the normal force as a function of tip-surface distance, which have been observed in the pull-off experiments with supramolecular bonds.\cite{Bennewitz1} In agreement with such experimental observations, the results of simulations exhibit two distinct regimes of forced rupture, where the maximum normal force is independent of the unloading rate at low driving, and it grows for higher unloading rates. The transitional unloading rate separating these regimes increases with the rate of bond rupture.

In lateral sliding, the model predicts a multiplicity of possible distinctive regimes of motion -- each corresponding to a specific molecular complex -- with increasing, decreasing, or constant friction force trends as a function of velocity. Our simulations demonstrate that the position of the maximum of the friction force as a function of velocity is determined by the ratio of the rates for bond formation and rupture. We have also shown that slide-hold-slide experiments may provide a way to directly estimate the rate of bond formation. Finally, our model predicts an enhancement of adhesion during sliding that should be a characteristic feature of long supramolecular bonds. The latter may lead to a novel ``anti-ageing'' effect, in which the number of connected bonds, and correspondingly the static friction force, decreases at larger hold times.

\small
\section*{Acknowledgment}
We acknowledge enlightening discussions with Dr.\ J.\ Blass and Dr.\ R.\ Bennewitz from Leibnitz Institute for New Materials, Saarbr\"ucken (Germany).
R.\ G.\ and A.\ V.\ acknowledge financial support from the ERC Advanced Grant 320796 - MODPHYSFRICT and from the COST Action MP1303 \emph{Understanding and Controlling Nano and Mesoscale Friction}. M.\ U.\ acknowledges the financial support of the Israel Science Foundation, Grant No.\ 1316/13. M.\ M.\ acknowledges financial support from a fellowship program for outstanding postdoctoral researchers from China and India in Israeli Universities.

\footnotesize{

}

\end{document}